\documentclass[12pt,a4paper]{article} 
\usepackage{graphicx}
\usepackage{cite}
\usepackage{a4wide} 

\makeatletter

\newcommand{\re}{\mathop{\mathrm{Re}}} 
\renewcommand\theenumi{(\@roman\c@enumi)}

\makeatother
\begin{document}

\title{\textbf{MODELLING OF $1/f$ NOISE BY SEQUENCES OF STOCHASTIC PULSES OF
DIFFERENT DURATION}}

\author{J.~Ruseckas, B.~Kaulakys and M.~Alaburda\\ \textit{\small Institute of
Theoretical Physics and Astronomy, Vilnius University, }\\ \textit{\small
A.~Go\v{s}tauto 12, LT-2600 Vilnius}}

\maketitle
\begin{abstract}
We present and analyze the simple analytically solvable model of $1/f$
noise, which can be relevant for the understanding of the origin,
main properties and parameter dependencies of the flicker noise. In
the model, the currents or signals represented as sequences of the
random pulses, which recurrence time intervals between transit times
of pulses are uncorrelated with the shape of the pulse, are analyzed.
It is shown that for the pulses of fixed area with random duration,
distributed uniformly in a wide interval, $1/f$ behavior of the power
spectrum of the signal or current in wide range of frequency may be
obtained.

\vskip \baselineskip


\end{abstract}

\section{Introduction}

The origin and omnipresence of $1/f$ noise is one of the oldest problems of the
contemporary physics. Since the first observation of the flicker noise in the
currents of electron tubes by Johnson \cite{johnson}, fluctuations of signals
and physical variables exhibiting behavior characterized by a power spectral
density diverging at low frequencies like $1/f$ have been observed in a wide
variety of systems \cite{dutta}. The widespread occurrence suggest that some
underlying mechanism might exist. However, a fully satisfactory explanation has
not yet been found and the general theory of $1/f$ noise is still an open
question.

A simple procedures of integration or differentiation of the convenient (white
noise, Brownian motion or so) fluctuating signals do not yield in the signal
exhibiting $1/f$ noise. There are no simple linear, even stochastic,
differential equations generating signals with $1/f$ noise. Therefore, $1/f$
noise is often modeled as the superposition of Lorentzian spectra with a wide
range of relaxation times \cite{mcwhorter}. Summation or integration of the
Lorentzians with the appropriate weights may yield $1/f$ noise \cite{kaulakys3}.

In many cases the physical processes can be represented by a sequence of random
pulses. Recently, considering signals and currents as consisting of pulses we
have shown \cite{kaulakys1,kaulakys2,kaulakys4} that the intrinsic origin of
$1/f$ noise may be a Brownian motion of the interevent time of the signal
pulses, similar to the Brownian fluctuations of the signal amplitude, resulting
in $1/f^2$ noise.

The model, proposed in \cite{kaulakys1,kaulakys2}, can be extended taking into
account finite duration of the pulse. The spectrum of the signal, consisting of
the pulse sequences which belong to the class of Markov process, was
investigated in \cite{heiden,shick}.

In this article we present a different model of pulses. We consider a signal
consisting of a sequence of uncorrelated pulses. The shape of the pulses is
determined by only one random parameter --- pulse duration. We will show that by
suitably choosing of the distribution of the pulse duration the $1/f$ noise can
be obtained.

\section{Signal as sequence of pulses}

We will investigate a signal consisting from a sequence of pulses. We assume
that:
\begin{enumerate}
\item the pulse sequences are stationary and ergodic;
\item interevent times and the shapes of different pulses are independent.
\end{enumerate}
The general form of such signal can be written as
\begin{equation}
I(t)=\sum_kA_k(t-t_k)\label{eq:signal}
\end{equation}
where functions $A_k(t)$ determine the shape of individual pulses and time
moments $t_k$ determine when a pulse occurs. The power spectrum is given by the
equation
\begin{equation}
S(f)=\lim_{T\rightarrow\infty}\left\langle\frac{2}{T}\left|\int_{t_i}^{
t_f}I(t)e^{-i2\pi ft}dt\right|^2\right\rangle\label{eq:spectr}
\end{equation}
where $T=t_f-t_i$. Substituting Eq.~(\ref{eq:signal}) into Eq.~(\ref{eq:spectr})
we have
\begin{equation}
S(\omega)=\lim_{T\rightarrow\infty}\left\langle\frac{2}{T}\sum_{k,k'}e^{
i\omega(t_k-t_{k'})}\int_{t_i-t_k}^{t_f-t_k}du\int_{t_i-t_{k'}}^{t_f-t_{
k'}}du'A_k(u)A_{k'}(u')e^{i\omega(u-u')}\right\rangle\,.\label{eq:2}
\end{equation}
 We assume that functions $A_k(u)$ decrease sufficiently fast when
$|u|\rightarrow\infty$. Since $T\rightarrow\infty$, the bounds of the
integration in Eq.~(\ref{eq:2}) can be changed to $\pm\infty$ . We also assume
that time moments $t_k$ are not correlated with the shape of the pulse $A_k$.
Then the power spectrum is
\[
S(\omega)=\lim_{T\rightarrow\infty}\frac{2}{T}\sum_{k,k'}\left\langle e^{
i\omega(t_k-t_{k'})}\right\rangle\left\langle\int_{-\infty}^{+\infty}du\int_{
-\infty}^{+\infty}du'A_k(u)A_{k'}(u')e^{i\omega(u-u')}\right\rangle .
\]

After introducing the functions
\begin{equation}
\Psi_{k,k'}(\omega)=\left\langle\int_{-\infty}^{+\infty}duA_k(u)e^{i\omega
 u}\int_{-\infty}^{+\infty}du'A_{k'}(u')e^{-i\omega u'}\right\rangle
\end{equation}
and
\begin{equation}
\chi_{k,k'}(\omega)=\left\langle e^{i\omega(t_k-t_{k'})}\right\rangle
\end{equation}
 the spectrum can be written as
\begin{equation}
S(\omega)=\lim_{T\rightarrow\infty}\frac{2}{T}\sum_{k,k'}\chi_{
k,k'}(\omega)\Psi_{k,k'}(\omega).\label{eq:spectr2}
\end{equation}

\subsection{Stationary process}

Equation (\ref{eq:spectr2}) can be further simplified assuming that the process
is stationary. In the stationary case all averages can depend only on $k-k'$.
Then
\begin{equation}
\Psi_{k,k'}(\omega)\equiv\Psi_{k-k'}(\omega).
\end{equation}
and
\begin{equation}
\chi_{k,k'}(\omega)\equiv\chi_{k-k'}(\omega).
\end{equation}
Equation (\ref{eq:spectr2}) then reads
\[
S(\omega)=\lim_{T\rightarrow\infty}\frac{2}{T}\sum_{k,k'}\chi_{k
-k'}(\omega)\Psi_{k-k'}(\omega).
\]
Changing the variables into $k\equiv k'$ and $q\equiv k-k'$ and changing the
order of summation we obtain
\begin{eqnarray*}
S(\omega)& = &\lim_{T\rightarrow\infty}\frac{2}{T}\sum_{q=1}^{k_{\mathrm{max}}
-k_{\mathrm{min}}}\sum_{k=k_{\mathrm{min}}}^{k_{\mathrm{max}}
-q}\chi_q(\omega)\Psi_q(\omega)\\
 &  & +\lim_{T\rightarrow\infty}\frac{2}{T}\sum_{q=k_{\mathrm{min}}-k_{\mathrm{
max}}}^{-1}\sum_{k=k_{\mathrm{min}}-q}^{k_{\mathrm{
max}}}\chi_q(\omega)\Psi_q(\omega)\\
 &  & +\lim_{T\rightarrow\infty}\frac{2}{T}\sum_{k=k_{\mathrm{min}}}^{k_{
\mathrm{max}}}\Psi_0(\omega).
\end{eqnarray*}
Introducing $N=k_{\mathrm{max}}-k_{\mathrm{min}}$ we have
\begin{equation}
S(\omega)=2\Psi_0(\omega)\bar{\nu}+\lim_{T\rightarrow\infty}4\sum_{
q=1}^N\left(\bar{\nu}-\frac{q}{T}\right)\re\chi_q(\omega)\Psi_q(\omega)
\label{eq:9}
\end{equation}
where
\begin{equation}
\bar{\nu}=\lim_{T\rightarrow\infty}\left\langle\frac{N+1}{T}\right\rangle
\end{equation}
is the mean number of pulses per unit time.

When the sum $\sum_{q=1}^Nq\re\chi_q(\omega)\Psi_q(\omega)$ converges and
$T\rightarrow\infty$ then the second term in the sum (\ref{eq:9}) vanishes and
the spectrum is
\begin{eqnarray}
S(\omega)& = & 2\bar{\nu}\Psi_0(\omega)+4\bar{\nu}\sum_{q=1}^{
\infty}\re\chi_q(\omega)\Psi_q(\omega)\label{eq:s2}\\
 & = & 2\bar{\nu}\sum_{q=-\infty}^{\infty}\chi_q(\omega)\Psi_q(\omega).
\end{eqnarray}

\subsection{Fixed shape pulses}

When the shape of the pulses is fixed ($k$-independent) then the function
$\Psi_{k,k'}(\omega)$ does not depend on $k$ and $k'$ and, therefore,
$\Psi_{k,k'}(\omega)=\Psi_{0,0}(\omega)$. Then equation (\ref{eq:spectr2})
yields the power spectrum
\begin{equation}
S(\omega)=\Psi_{0,0}(\omega)\lim_{T\rightarrow\infty}\frac{2}{T}\sum_{
k,k'}\chi_{k,k'}(\omega)\equiv\Psi_{0,0}(\omega)S_{\delta}(\omega).
\end{equation}
This is the spectrum of one pulse multiplied by the spectrum of the sequence of
$\delta$-shaped pulses $S_{\delta}(\omega)$. It has been shown
\cite{kaulakys1,kaulakys2} that the spectrum of such a sequence can exhibit
$1/f$-like behaviour in a broad frequency range if the interevent times
$\tau_k=t_k-t_{k-1}$ follow an autoregressive process.

\subsection{Uncorrelated pulses}

When the pulses are uncorrelated and $k\neq k'$ then
\begin{eqnarray*}
\Psi_{k-k'}(\omega)& = &\left\langle\int_{-\infty}^{+\infty}A_k(u)e^{i\omega
 u}du\right\rangle\left\langle\int_{-\infty}^{+\infty}A_{k'}(u')e^{-i\omega
 u'}du'\right\rangle\\
 & = & |\langle F_k(\omega)\rangle|^2
\end{eqnarray*}
where
\begin{equation}
F_k(\omega)=\int_{-\infty}^{+\infty}A_k(u)e^{i\omega u}du\,.
\end{equation}
is the Fourier transform of the pulse $A_k$. When $k=k'$ then
\[
\Psi_0(\omega)=\left\langle |F_k(\omega)|^2\right\rangle .
\]
>From Eq.~(\ref{eq:s2}) we obtain the spectrum
\begin{equation}
S(\omega)=2\bar{\nu}\left\langle |F_k(\omega)|^2\right\rangle
+4\bar{\nu}|\langle F_k(\omega)\rangle|^2\sum_{q=1}^{
\infty}\re\chi_q(\omega).\label{eq:s3}
\end{equation}

When the interevent times $\tau_k=t_k-t_{k-1}$ are random and uncorrelated then
\begin{equation}
\chi_q(\omega)=\left\langle e^{i\omega(t_{k+q}-t_k)}\right\rangle =\left\langle
 e^{i\omega\tau_k}\right\rangle^q\equiv\chi_{\tau}(\omega)^q.
\end{equation}
>From Eq.~(\ref{eq:s3}) we obtain
\begin{equation}
S(\omega)=2\bar{\nu}\left\langle |F_k(\omega)|^2\right\rangle +4\bar{
\nu}|\langle F_k(\omega)\rangle|^2\re\frac{\chi_{\tau}(\omega)}{1-\chi_{
\tau}(\omega)}.
\end{equation}
Here
\begin{equation}
\bar{\nu}=\left[-i\left.\frac{d\chi_{\tau}(\omega)}{d\omega}\right|_{
\omega=0}\right]^{-1}.
\end{equation}

If the occurrence times of the pulses $t_k$ are distributed according to Poisson
process then the interevent time probability distribution is
$\Psi(\tau)=\frac{1}{\bar{\tau}}e^{-\frac{\tau}{\bar{\tau}}}$. The
characteristic function obeys the equality
$\mathrm{Re}\frac{\chi_{\tau}(\omega)}{1-\chi_{\tau}(\omega)}=0$ and the
spectrum is
\begin{equation}
S(\omega)=2\bar{\nu}\left\langle |F_k(\omega)|^2\right\rangle .
\label{eq:uncorr2}
\end{equation}
We will investigate this case more deeply.

\section{Pulses of variable duration }

Let the only random parameter of the pulse is the duration. We take the form of
the pulse as
\begin{equation}
A_k(t)=T_k^{\beta}A\left(\frac{t}{T_k}\right),\label{eq:pulse}
\end{equation}
where $T_k$ is the characteristic duration of the pulse. The value $\beta=0$
corresponds to fixed height pulses; $\beta=-1$ corresponds to constant area
pulses. Differentiating the fixed area pulses we obtain $\beta=-2$. The Fourier
transform of the pulse (\ref{eq:pulse}) is
\[
F_k(\omega)=\int_{-\infty}^{+\infty}T_k^{\beta}A\left(\frac{t}{T_k}\right)e^{
i\omega t}dt=T_k^{\beta+1}\int_{-\infty}^{+\infty}A(u)e^{i\omega T_ku}du\equiv
 T_k^{\beta+1}F(\omega T_k).
\]
>From Eq.~(\ref{eq:uncorr2}) the power spectrum is
\begin{equation}
S(\omega)=2\bar{\nu}\left\langle T_k^{2\beta+2}|F(\omega T_k)|^2\right\rangle .
\end{equation}
Introducing the probability density $P(T_k)$ of the pulses durations $T_k$ we
can write
\begin{equation}
S(\omega)=2\bar{\nu}\int_0^{\infty}T_k^{2\beta+2}|F(\omega
 T_k)|^2P(T_k)dT_k\,.\label{eq:spektr3}
\end{equation}
 If $P(T_k)$ is a power-law distribution, then the expressions for the spectrum
are similar for all $\beta$.

\subsection{Spectrum at small frequencies $\omega$}

For small frequencies we expand the Fourier transform of the pulse into Taylor
series. The first coefficients are
\begin{equation}
F(0)=a,\quad\frac{dF(0)}{d\omega}=ia\langle t\rangle,\quad\frac{d^2F(0)}{
d\omega^2}=-a\langle t^2\rangle,
\end{equation}
where
\begin{equation}
a=\int_{-\infty}^{+\infty}A(t)dt
\end{equation}
is the area of the pulse,
\begin{equation}
\langle t\rangle=\frac{1}{a}\int_{-\infty}^{+\infty}tA(t)dt,\quad\langle
 t^2\rangle=\frac{1}{a}\int_{-\infty}^{+\infty}t^2A(t)dt.
\end{equation}
Then the spectrum from Eq.~(\ref{eq:spektr3}) is
\[
S(\omega)\approx2\bar{\nu}a^2\int_0^{\infty}T_k^{2\beta+2}(1-\Delta
 t^2\omega^2T_k^2)P(T_k)dT_k\,,
\]
where $\Delta t^2=\langle t^2\rangle-\langle t\rangle^2$. We obtain
\begin{equation}
S(\omega)=2\bar{\nu}a^2\langle T_k^{2\beta+2}\rangle(1-\Delta t^2\omega^2\langle
 T_k^{2\beta+4}\rangle),
\end{equation}
where
\begin{equation}
\langle T_k^{\xi}\rangle=\int_0^{\infty}T_k^{\xi}P(T_k)dT_k\,.
\end{equation}

\subsection{Power-law distribution}

We take the power-law distribution of pulse durations
\begin{equation}
P(T_k)=\left\{
\begin{array}{ll}
\frac{\alpha+1}{T_{\mathrm{max}}^{\alpha+1}-T_{\mathrm{min}}^{\alpha+1}}T_k^{
\alpha}, & T_{\mathrm{min}}\leq T_k\leq T_{\mathrm{max}},\\
0, &\textrm{othervise}.\end{array}\right.\label{eq:pow}
\end{equation}
>From Eq.~(\ref{eq:spektr3}) we have the spectrum
\begin{eqnarray*}
S(\omega)& = & 2\bar{\nu}\frac{\alpha+1}{T_{\mathrm{max}}^{\alpha+1}-T_{\mathrm{
min}}^{\alpha+1}}\int_0^{\infty}T_k^{\alpha+2\beta+2}|F(\omega T_k)|^2dT_k\\
 & = &\frac{2\bar{\nu}(\alpha+1)}{\omega^{\alpha+2\beta+3}(T_{\mathrm{max}}^{
\alpha+1}-T_{\mathrm{min}}^{\alpha+1})}\int_{\omega T_{\mathrm{min}}}^{\omega
 T_{\mathrm{max}}}u^{\alpha+2\beta+2}|F(u)|^2du\,.
\end{eqnarray*}
When $\alpha>-1$ and
$\frac{1}{T_{\mathrm{max}}}\ll\omega\ll\frac{1}{T_{\mathrm{min}}}$ then the
expression for the spectrum can be approximated as
\begin{equation}
S(\omega)\approx\frac{2\bar{\nu}(\alpha+1)}{\omega^{\alpha+2\beta+3}(T_{\mathrm{
max}}^{\alpha+1}-T_{\mathrm{min}}^{\alpha+1})}\int_0^{\infty}u^{\alpha+2\beta
+2}|F(u)|^2du
\end{equation}

If $\alpha+2\beta+2=0$ then in the frequency domain
$\frac{1}{T_{\mathrm{max}}}\ll\omega\ll\frac{1}{T_{\mathrm{min}}}$ the spectrum
is
\begin{equation}
S(\omega)\approx\frac{2\bar{\nu}(\alpha+1)}{\omega(T_{\mathrm{max}}^{\alpha+1}
-T_{\mathrm{min}}^{\alpha+1})}\int_0^{\infty}|F(u)|^2du\,.\label{eq:result}
\end{equation}
We obtained $1/f$ spectrum. The condition $\alpha+2\beta+2=0$ is satisfied,
e.g., for the fixed area pulses ($\beta=-1$) and uniform distribution of pulse
durations or for fixed height pulses ($\beta=0$) and uniform distribution of
inverse durations $\gamma=T_k^{-1}$ , i.e. for $P(T_k)\propto T_k^{-2}$ .

If $\alpha+2\beta+4=0$ then in the frequency domain
$\frac{1}{T_{\mathrm{max}}}\ll\omega\ll\frac{1}{T_{\mathrm{min}}}$ the spectrum
is
\begin{equation}
S(\omega)\approx\frac{2\bar{\nu}(\alpha+1)\omega}{(T_{\mathrm{max}}^{\alpha+1}
-T_{\mathrm{min}}^{\alpha+1})}\int_0^{\infty}|F(u)|^2\frac{du}{u^2}\,.
\end{equation}
Such a spectrum can be obtained after differentiation of the signal exhibiting
$1/f$ spectrum.

\section{Example}

\subsection{Rectangular pulses}

As an example we will obtain the spectrum of rectangular constant area pulses.
The duration of the pulse is $T_k$. The Fourier transform of the pulse is
\begin{equation}
F(\omega T_k)=a\int_0^1du\, e^{i\omega T_ku}=a\frac{e^{i\omega T_k}-1}{i\omega
 T_k}=ae^{i\frac{\omega T_k}{2}}\frac{2\sin\left(\frac{\omega T_k}{2}\right)}{
\omega T_k}.\label{eq:four1}
\end{equation}
Then the spectrum according to Eqs.~(\ref{eq:spektr3}), (\ref{eq:pow}) and
(\ref{eq:four1}) is
\begin{eqnarray}
S(\omega)& = &\frac{4\bar{\nu}a^2(\alpha+1)(T_{\mathrm{max}}^{\alpha-1}-T_{
\mathrm{min}}^{\alpha-1})}{\omega^2(\alpha-1)(T_{\mathrm{max}}^{\alpha+1}-T_{
\mathrm{min}}^{\alpha+1})}+\frac{4\bar{\nu}a^2(\alpha+1)}{\omega^{\alpha+1}(T_{
\mathrm{max}}^{\alpha+1}-T_{\mathrm{min}}^{\alpha+1})}\nonumber\\
 &  &\times\re\left\{i^{1-\alpha}\left(\Gamma(\alpha-1,i\omega T_{\mathrm{max}})
-\Gamma(\alpha-1,i\omega T_{\mathrm{min}})\right)\right\},
\end{eqnarray}
where $\Gamma(a,z)$ is the incomplete gamma function,
$\Gamma(a,z)=\int_z^{\infty}u^{a-1}e^{-u}du$.

When $-1<\alpha<1$ then the term with $\Gamma(\alpha-1,i\omega
T_{\mathrm{max}})$ is small and can be neglected. We also assume that
$T_{\mathrm{min}}\ll T_{\mathrm{max}}$ and when $\alpha<-1$ we neglect the term
$\left(\frac{T_{\mathrm{min}}}{T_{\mathrm{max}}}\right)^{\alpha+1}$. Then we
have
\begin{equation}
S(\omega)\approx-\frac{4\bar{\nu}a^2(\alpha+1)}{\omega^{\alpha+1}T_{\mathrm{
max}}^{\alpha+1}}\cos\left(\frac{\pi}{2}(\alpha-1)\right)\Gamma(\alpha-1).
\end{equation}
For $\alpha=0$ we have the uniform distribution of the pulses duration. Using
the result of the limit
\begin{equation}
\lim_{\alpha\rightarrow0}\cos\left(\frac{\pi}{2}(\alpha-1)\right)\Gamma(\alpha
-1)=-\frac{\pi}{2}\,,
\end{equation}
we obtain $1/f$ spectrum
\begin{equation}
S(\omega)\approx\frac{2\pi\bar{\nu}a^2}{\omega T_{\mathrm{max}}}.
\label{eq:rect1f}
\end{equation}

\begin{figure}
\includegraphics[width=0.70\textwidth]{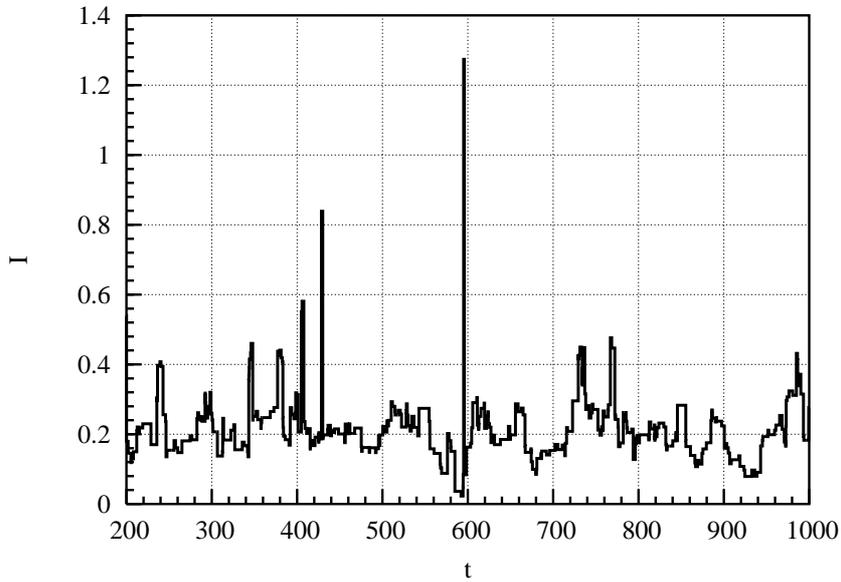}
\caption{Typical signal consisting from the fixed area rectangular pulses with
uniformly distributed durations. The time intervals between the pulses are
distributed according to Poisson process with the average $\bar{\tau}=5$. The
used parameters are $T_{\mathrm{min}}=0.01$, $T_{\mathrm{max}}=100$ .}
\label{fig:signal}
\end{figure}

\begin{figure}
\includegraphics[width=0.70\textwidth]{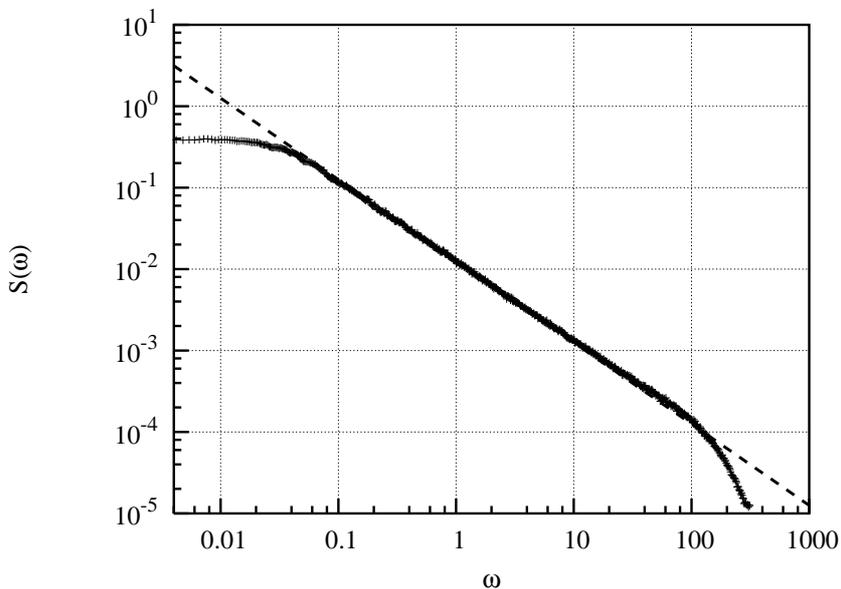}
\caption{The spectrum of the signal consisting from the fixed area pulses with
uniformly distributed durations. The used parameters are the same as in
Fig.~\ref{fig:signal}. The dashed line corresponds to the spectrum obtained
according Eq.~(\ref{eq:rect1f})}
\label{fig:1fspectr}
\end{figure}

The spectrum was also obtained from numerical calculations. Typical signal for
rectangular fixed area pulses is shown in Fig.~\ref{fig:signal} and the power
spectrum in Fig.~\ref{fig:1fspectr}.

\subsection{Differentiated rectangular pulses}

\begin{figure}
\includegraphics[width=0.70\textwidth]{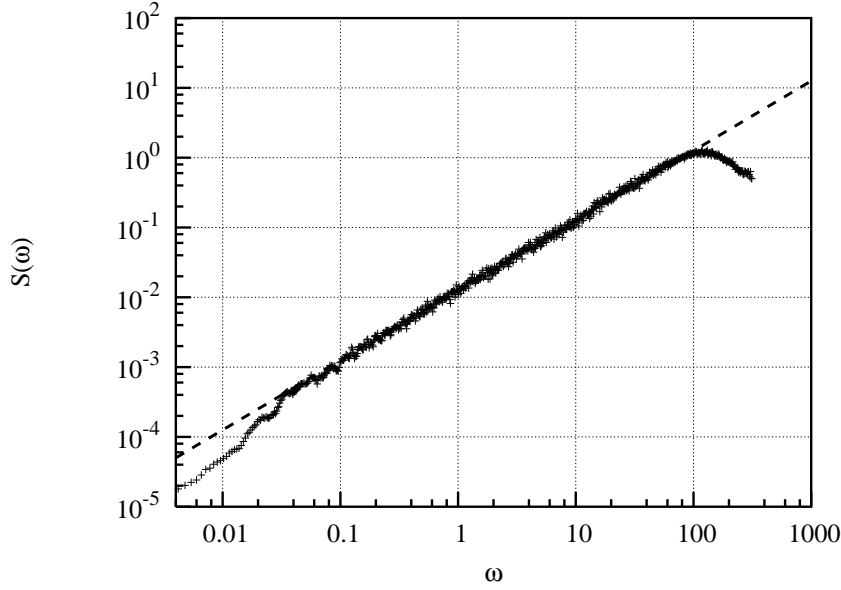}
\caption{The spectrum of differentiated signal consisting from the fixed area
pulses. The used parameters are the same as in Fig.~\ref{fig:signal}. The dashed
line corresponds to the spectrum obtained according Eq.~(\ref{eq:frect}).}
\label{fig:fspectr}
\end{figure}

After differentiation of the signal the shape of the pulse becomes
\begin{equation}
A\left(\frac{t}{T_k}\right)=a\left(\delta\left(\frac{t}{T_k}\right)
-\delta\left(\frac{t}{T_k}-1\right)\right).
\end{equation}
The Fourier transform of the pulse is
\begin{equation}
F(\omega T_k)=a(1-e^{i\omega T_k})=-2iae^{i\frac{\omega T_k}{2}}\sin\left(\frac{
\omega T_k}{2}\right).
\end{equation}
Then the spectrum is
\begin{eqnarray}
S(\omega)& = &\frac{4\bar{\nu}a^2(\alpha+1)(T_{\mathrm{max}}^{\alpha-1}-T_{
\mathrm{min}}^{\alpha-1})}{(\alpha-1)(T_{\mathrm{max}}^{\alpha+1}-T_{\mathrm{
min}}^{\alpha+1})}+\frac{4\bar{\nu}a^2(\alpha+1)}{\omega^{\alpha-1}(T_{\mathrm{
max}}^{\alpha+1}-T_{\mathrm{min}}^{\alpha+1})}\nonumber\\
 &  &\times\re\left(i^{1-\alpha}(\Gamma(\alpha-1,i\omega T_{\mathrm{max}})
-\Gamma(\alpha-1,i\omega T_{\mathrm{min}}))\right).
\end{eqnarray}
When $\alpha=0$ we obtain
\begin{equation}
S(\omega)\approx\frac{2\pi\bar{\nu}a^2}{T_{\mathrm{max}}}\omega.\label{eq:frect}
\end{equation}
The spectrum, obtained numerically, is shown in Fig.~\ref{fig:fspectr}.

\section{Conclusion}

We investigate signals consisting of a sequence of uncorrelated pulses with
random durations. By suitably choosing distribution of the pulse duration the
$1/f$ noise can be obtained. Signal of fixed area pulses yields $1/f$ noise when
width of the pulse is uniformly distributed in a wide interval. The spectrum is
given by Eq.~(\ref{eq:result}). This conclusion does not depend on particular
shape of the pulse. For the fixed amplitude pulses $1/f$ spectrum yields when
the inverse duration of the pulses $\gamma=T_k^{-1}$ is distributed uniformly or
$P(T_k)\propto T_k^{-2}$

\end{document}